\documentclass{mn2e}
\input{epsf}

\title[Adventures of a tidally induced bar]
{Adventures of a tidally induced bar}

\author[E. L. {\L}okas et al.]
    {E. L. {\L}okas,$^{1}$ E. Athanassoula,$^{2}$ V. P. Debattista,$^{3}$ M. Valluri,$^{4}$ A. del Pino,$^{5}$
     \newauthor M. Semczuk,$^{6}$ G. Gajda$^{6}$ and K. Kowalczyk$^{1,7}$
    \\
    \\
    $^1$Nicolaus Copernicus Astronomical Center, Bartycka 18, 00-716 Warsaw, Poland; lokas@camk.edu.pl\\
    $^2$Laboratoire d'Astrophysique de Marseille (LAM), UMR6110, CNRS/Universit\'e de Provence,
    Technop\^{o}le de Marseille Etoile,\\ 38 rue Fr\'ed\'eric Joliot Curie, F-13388 Marseille C\'edex 13, France\\
    $^3$Jeremiah Horrocks Institute, University of Central Lancashire, Preston PR1 2HE, UK\\
    $^4$Department of Astronomy, University of Michigan, Ann Arbor, MI 48109, USA\\
    $^5$Instituto de Astrof\'{i}sica de Canarias, Calle V\'{i}a L\'{a}ctea s/n, E-38200 La Laguna, Tenerife,
	Canary Islands, Spain\\
    $^6$Astronomical Observatory of the Jagiellonian University, Orla 171, 30-244 Cracow, Poland \\
    $^7$Warsaw University Observatory, Al. Ujazdowskie 4, 00-478 Warsaw, Poland}

\begin{document}

\maketitle

\begin{abstract}
Using $N$-body simulations, we study the properties of a bar induced in a disky dwarf galaxy as a
result of tidal interaction with the Milky Way. The bar forms at the first pericentre passage and
survives until the end of the evolution at 10 Gyr. Fourier decomposition of the bar reveals that
only even modes are significant and preserve a hierarchy so that the bar mode is always the
strongest. They show a characteristic profile with a maximum, similar to simulated bars forming in
isolated galaxies and observed bars in real galaxies. We adopt the maximum of the bar mode as a
measure of the bar strength and we estimate the bar length by comparing the density profiles along
the bar and perpendicular to it. The bar strength and the bar length decrease with time, mainly at
pericentres, as a result of tidal torques acting at those times and not to secular evolution. The
pattern speed of the bar varies significantly on a time scale of 1 Gyr and is controlled by the
orientation of the tidal torque from the Milky Way. The bar is never tidally locked, but we discover
a hint of a 5/2 orbital resonance between the third and fourth pericentre passage. The speed of the
bar decreases in the long run so that the bar changes from initially rather fast to slow in the
later stages. The boxy/peanut shape is present for some time and its occurrence is preceded by a
short period of buckling instability.
\end{abstract}

\section{Introduction}

Observations indicate that a significant fraction of spiral galaxies are barred (e.g. Cheung et al. 2013
and references therein). While precise measurements of the
bar fraction vary (between 20 and 70 percent, depending on the sample) there is no doubt that the phenomenon is a
rather common feature in galaxies. In addition, comparisons of the frequency of bar presence in nearby galaxies
to the one in high-redshift objects lead to the conclusion that bars occur relatively late in the evolution of
galaxies (Sheth et al. 2008).

The essence of bar formation is the transformation of circular orbits in the stellar disk into elongated ones
(for reviews on dynamics of bars see e.g. Sellwood \& Wilkinson 1993; Athanassoula 2013).
While a simplified analytic description of the phenomenon is possible (Binney \& Tremaine 2008), the problem can be
fully tackled only using $N$-body simulations. This approach was pioneered by Miller and Smith (1979) who studied
the bar evolution and discussed the bar pattern speed, the particle orbits and the predicted observational properties
of bars. Sparke \& Sellwood (1987) studied the bar formation via disk instability, performed detailed classification
of stellar orbits in the bar and found the bar to be stable and robust. While such instability is indeed generally
believed to be responsible for bar formation, we still do not have a full grasp of all the intricacies
of bar formation and evolution and alternative ways to form a bar have been considered.

Combes et al. (1990) were among the first to discuss the box and peanut shapes generated by stellar bars and Raha
et al. (1991) discovered a mechanism that may be responsible for these shapes in the form of the buckling instability.
The instability drives the stars out of the galactic plane and may significantly weaken the bar (for more
recent developments see e.g. Athanassoula 2005; Martinez-Valpuesta, Shlosman \& Heller 2006;
Saha, Pfenniger \& Taam 2013). Insight into its nature can be obtained via studies of vertical
orbital instabilities (e.g. Binney 1978; Pfenniger 1984; Skokos, Patsis \& Athanassoula 2002a,b).

Additional complication in the study of bar dynamics is introduced by the presence of extended dark matter
haloes surrounding the disks (Debattista \& Sellwood 2000; Athanassoula 2002, 2003;
Dubinski, Berentzen \& Shlosman 2009; Saha \& Naab 2013). Overall, the evolution of the bar's major properties,
such as its strength or pattern speed
depends on a plethora of parameters and has been only partially explored in simulations (Athanassoula \& Misiriotis
2002; Klypin et al. 2009; Athanassoula, Machado \& Rodionov 2013).

One factor that has been relatively underexplored is the effect of galaxy interactions.
Previous studies of the influence of interactions on the properties of the bars discussed mainly the effect of a
satellite on the bar in the normal-size galaxy (Gerin, Combes \& Athanassoula 1990; Sundin, Donner \& Sundelius 1993;
Miwa \& Noguchi 1998; Mayer \& Wadsley 2004; Romano-D\'{i}az et al. 2008).
In this paper we are interested in a different phenomenon, that of a tidally induced
bar in a dwarf galaxy interacting with a bigger host. Important hints that such a process may be important in
shaping the properties of present-day dwarfs are provided by observations that bars form later in fainter
galaxies (Sheth et al. 2008) and that bar fraction is higher in fainter objects in dense environments such as
galaxy clusters (Lansbury, Lucey \& Smith 2014). A possible interpretation of these results is that dwarf galaxies
are born with dynamically hotter disks which delays bar formation (Athanassoula \& Sellwood 1986; Sheth et al. 2012)
until they are accreted by a more massive galaxy or become a member of a group or cluster where they are affected
by tidal forces. An example of such a tidally induced bar could be the one in M82, resulting from an interaction
with M81 (Wills et al. 2000).

The formation of tidally induced bars in dwarf galaxies has been studied mainly in the context of the tidal stirring
scenario for the formation of dwarf spheroidal (dSph) galaxies in the Local Group (Mayer et al. 2001; Klimentowski
et al. 2009; Kazantzidis et al. 2011; {\L}okas, Kazantzidis \& Mayer 2011).
The simulations of this process revealed that the formation
of a bar is a natural intermediate stage of a dwarf galaxy in its morphological evolution from a disk toward a
spheroid that occurs for a variety of dwarf's orbits around the host and for different initial structural parameters.
{\L}okas et al. (2012) measured the shapes of simulated dSph galaxies and compared them to the shapes of classical
dSph satellites of the Milky Way quantified in the same way. Bar-like surface density distributions were found in the
Sagittarius, Ursa Minor and possibly Carina dwarfs. Elongated shapes, suggestive of the presence of the bar are also
seen in the recently discovered ultra-faint dwarfs like Hercules (Coleman et al. 2007) and Ursa Major II
(Mu\~noz, Geha \& Willman 2010). Note that although the Large Magellanic Cloud (LMC) is also known to possess a bar,
this structure is probably not of tidal origin. Indeed, according to the most probable scenario the
LMC is at its first pericentre around the Milky Way (Besla et al. 2007) and its pericentric distance is too large
(50 kpc) for the tidal forces to be effective.

The Sagittarius dwarf seems to be the most obvious candidate for a barred galaxy among the Local Group dSphs.
{\L}okas et al. (2010) proposed an evolutionary model of this dwarf starting from a disk embedded in a dark matter
halo. After the first pericentre passage the stellar component of the dwarf transforms into a bar that survives until
the next pericentre which was identified as the present stage of the dwarf we observe. The shape of the dwarf at this
time matches very well the actual elongated shape determined from observations by Majewski et al. (2003). The model
also explains the lack of rotation signal in the data (Frinchaboy et al. 2012).

In this work we look in more detail into the properties of a tidally induced bar on a typical orbit around the Milky Way.
In section 2 we present the simulation used for this study. In section 3 we describe the evolution of the dwarf
galaxy using global measurements of its properties. Sections 4 and 5 focus on the strength and length of the bar,
while section 6 is devoted to the pattern speed of the bar and its interpretation. The discussion follows in
section 7.

\section{The simulation}

Our simulation setup consisted of live models of two galaxies: the Milky Way-like host and the dwarf galaxy.
The $N$-body realizations of both galaxies were generated via procedures described in Widrow \& Dubinski (2005)
and Widrow, Pym \& Dubinski (2008). The procedures allow us to generate near-equilibrium models of galaxies composed
of a disk, a bulge and a halo with required properties. Both our galaxies contained exponential disks embedded in
NFW (Navarro, Frenk \& White 1997) dark matter haloes. The dark haloes were smoothly truncated at the radius
close to the virial radius in order to make their masses finite. Each component of each galaxy contained
$10^6$ particles ($4 \times 10^6$ particles total).

The dwarf galaxy model was similar to the default model
used in recent simulations of tidal stirring (Kazantzidis et al. 2011; {\L}okas et al. 2011). The dark
halo of the dwarf had a mass $M_{\rm h} = 10^9$ M$_{\odot}$ and concentration $c=20$. The disk had a
mass $M_{\rm d} = 2 \times 10^7$ M$_{\odot}$, exponential scale-length $R_{\rm d} = 0.41$ kpc and thickness
$z_{\rm d}/R_{\rm d} = 0.2$. The coldness of the disk is controlled by the central radial velocity dispersion
which we assume to be $\sigma_{R0} = 10$ km s$^{-1}$. This translates to the Toomre parameter $Q = 3.82$ at
$R = 2.5 R_{\rm d}$ and guarantees that our dwarf is stable against formation of the bar in isolation for the
time scales of interest here. We verified this by evolving the dwarf galaxy in isolation for 10 Gyr.

The host galaxy was chosen to resemble the model MWb of Widrow \& Dubinski (2005). It had a dark matter halo of mass
$M_{\rm H} = 7.7 \times 10^{11}$ M$_{\odot}$ and concentration $c=27$. The disk of the host had a mass $M_{\rm D}
= 3.4 \times 10^{10}$ M$_{\odot}$, the scale-length $R_{\rm D} = 2.82$ kpc and thickness $z_{\rm D} = 0.44$ kpc.
The central radial velocity dispersion of the disk was $\sigma_{R0} = 121$ km s$^{-1}$ which corresponds to
the Toomre parameter $Q = 2.2$ again making the disk stable against bar formation for time scales of interest.

Although there is evidence that the Milky Way has a bar (e.g. Blitz \& Spergel 1991; Dwek et al. 1995;
Martinez-Valpuesta \& Gerhard 2011; Romero-G\'omez et al. 2011), we specifically chose a model for the host
galaxy that does not form a bar since this makes the host potential constant in time. The main reason for
this choice is that we are most interested here in modeling the effect of the tidal forces on the growth
and evolution of the bar in the dwarf. The presence of a time dependent bar  in the Milky Way-like host
would induce a second time dependence whose influence would be difficult to disentangle from the tidal
effects. Furthermore, both observational (e.g. Stanek et al. 1997) and theoretical (e.g. Shen et al. 2010)
studies indicate that the bar in the Milky Way is not very strong. Moreover, its half-length is of the order of 4 kpc,
i.e. much smaller than our adopted pericentre distance. It is thus unlikely to significantly influence the evolution
of the tidal bar in the dwarf.
For simplicity we also neglect other components of the Milky Way structure
such as the bulge, the stellar halo, the distinction into thin/thick disk etc. The mass of these components is at least a
few times smaller than the disk mass we assume and probably of the order of the uncertainty of the mass distribution
in the two main components (Widrow \& Dubinski 2005).

We placed the dwarf galaxy initially at an apocentre of a typical, eccentric orbit around the Milky Way
with apo- to pericentre distance ratio of $r_{\rm apo}/r_{\rm peri} = 120/25$ kpc. Due to a rather small mass
of the dwarf the orbit decays only a little in time as a result of dynamical friction. The dwarf's disk, the disk
of the Milky Way and the orbit were initially all coplanar and the dwarf's disk was in prograde rotation with
respect to the orbit. This configuration, together with the stability of the Milky Way's disk, makes sure that
the tidal forces experienced by the dwarf during the evolution are only controlled by the dwarf's distance
from the host galaxy and not by other subtle changes of the potential.

The evolution of the dwarf was followed for 10 Gyr using the GADGET-2 $N$-body code (Springel, Yoshida \& White 2001;
Springel 2005) and we saved 201 simulation outputs, one every 0.05 Gyr (which is significantly smaller
than the dynamical time of stars in the dwarf's disk).
The code configuration was that of Newtonian space with vacuum boundary conditions and the gravity
calculations were done with a tree option. We adopted the softening scales of $\epsilon_{\rm d} = 0.02$ kpc and
$\epsilon_{\rm h} = 0.06$ kpc for the disk and halo of the dwarf and $\epsilon_{\rm D} = 0.05$ kpc and
$\epsilon_{\rm H} = 2$ kpc for the disk and halo of the Milky Way, respectively. These choices were informed
by the study of Power et al. (2003) and allow to avoid strong discreteness and two-body effects in the case of
systems of given characteristic scales and particle numbers.

\section{Overview of the evolution}

The repeated action of the tidal force from the Milky Way made the dwarf galaxy evolve as envisioned by the tidal
stirring scenario originally proposed by Mayer et al. (2001) and studied in more detail by Klimentowski et al. (2009),
Kazantzidis et al. (2011) and {\L}okas et al. (2011, 2012). The main signatures of such evolution involve the mass
loss, the morphological transformation of the stellar component and the changes in the kinematics of the stars. Since
we are interested here in the formation and evolution of the bar we focus on the latter two. In order to give an idea
of the overall evolution of the dwarf we first perform rough measurements of the main, global features.

Inspection of the final state of the dwarf reveals that, in spite of the strong mass loss, it still possesses a well
visible bound stellar component of radius of the order of 1 kpc. Therefore we may measure the properties of the dwarf
at all times using stars within some fixed radius comparable to this value, or smaller.
One could choose to measure
the dwarf's properties at some characteristic scale such as the radius where maximum circular velocity occurs
(as was done in Klimentowski et al. 2009 and Kazantzidis et al. 2011), the half-light radius ({\L}okas et al.
2011, 2012) or the break radius (where the transition to the tidal tails occurs, see e.g.
{\L}okas, Gajda \& Kazantzidis 2013). However, the caveat of such measurements
is that such radii also evolve in time and the interpretation of the results may not be straightforward.

\begin{figure}
\begin{center}
    \leavevmode
    \epsfxsize=8cm
    \epsfbox[5 5 182 270]{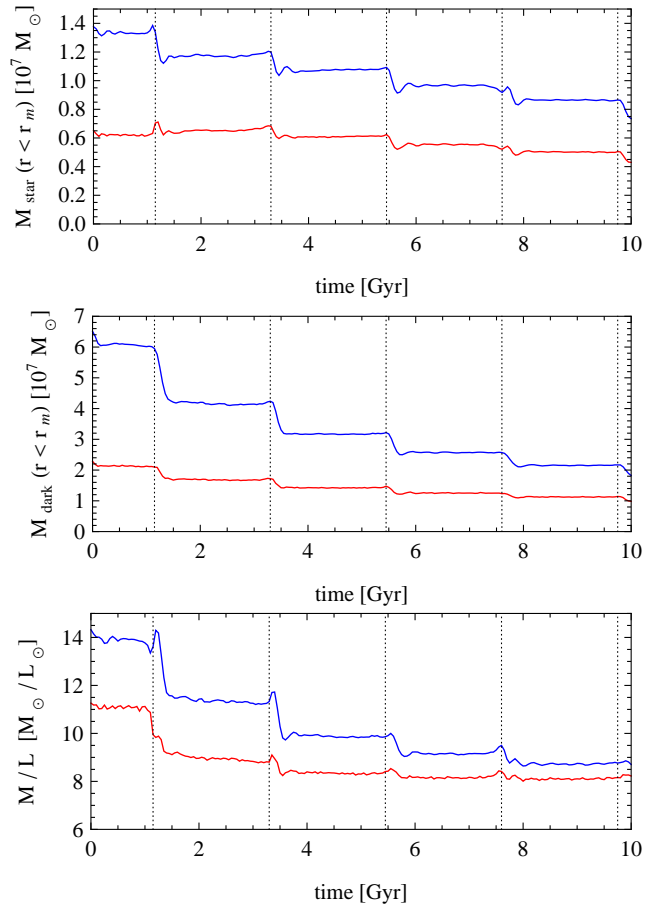}
\end{center}
\caption{The evolution of the mass of the dwarf galaxy. The upper panel shows measurements of the mass of stars,
the middle panel the mass of the dark matter component, and the lower one the mass-to-light ratio (assuming
$M/L = 2.5$ M$_{\odot}$/L$_{\odot}$ for the stars) within a fixed maximum radius $r_{\rm m}$.
In each panel the red curve corresponds to measurements within $r_{\rm m} = 0.5$ kpc and the blue one
within $r_{\rm m} = 1$ kpc. Vertical dotted lines indicate pericentre passages.}
\label{mass}
\end{figure}

\begin{figure}
\begin{center}
    \leavevmode
    \epsfxsize=8cm
    \epsfbox[5 5 182 182]{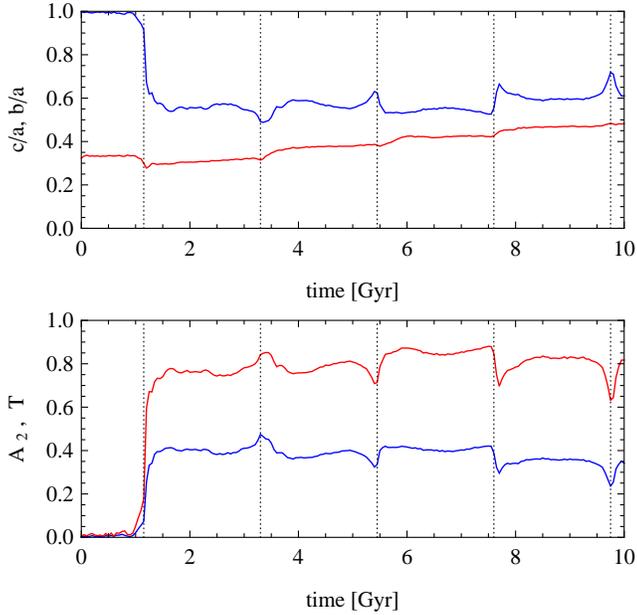}
\end{center}
\caption{The evolution of the shape of the stellar component of the dwarf galaxy. The upper panel shows
the evolution of the axis ratios $c/a$ (red lower line) and $b/a$ (blue upper line) in time. The lower panels shows
the evolution of the bar mode $A_2$ (blue lower line) and the triaxiality parameter $T$ (red upper line) in time.
All measurements were performed for stars within a constant radius of 0.5 kpc. Vertical dotted lines indicate
pericentre passages.}
\label{shape}
\end{figure}

\begin{figure}
\begin{center}
    \leavevmode
    \epsfxsize=8cm
    \epsfbox[5 5 182 270]{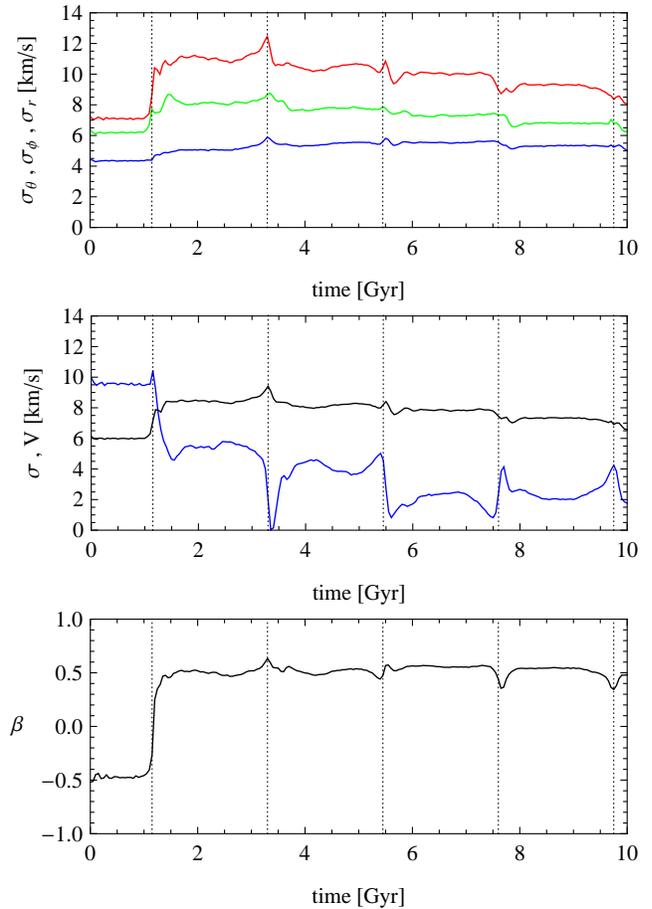}
\end{center}
\caption{The evolution of the kinematics of the stellar component of the dwarf galaxy. Upper panel:
the evolution of the velocity dispersions of the stars in spherical coordinates $\sigma_r$ (red upper line),
$\sigma_\phi$ (green middle line) and $\sigma_\theta$ (blue lower line). Middle panel:
the evolution of the averaged 1D velocity dispersion $\sigma$ (black line) and the rotation velocity $V$ (blue line).
Lower panel: evolution of the anisotropy parameter $\beta$.
All measurements were performed for stars within a constant radius of 0.5 kpc. Vertical dotted lines indicate
pericentre passages.}
\label{kinematics}
\end{figure}

For each simulation output we determine the centre of the dwarf galaxy by calculating the centre of mass of the stars
iteratively in decreasing radii until convergence is reached.
We start the analysis of the properties of the dwarf galaxy by measuring
the mass of stars and dark matter within a fixed radius $r_{\rm m}$ from the centre of the dwarf equal
to 1 and 0.5 kpc. The results are shown in the upper (stars) and middle
(dark matter) panel of Figure~\ref{mass} with the red curve corresponding to measurements within
$r_{\rm m} = 0.5$ kpc and the blue one within $r_{\rm m} = 1$ kpc. In the lower panel of the Figure we also plot
the mass-to-light ratio within these radii assuming $M/L = 2.5$ M$_{\odot}$/L$_{\odot}$ for the stars. The results
show that mass in both components is systematically lost from within both limiting radii, except for the period
after the first pericentre passage where the stellar content within 0.5 kpc is not diminished. The mass-to-light
ratio systematically decreases to converge to about $(8-9)$ M$_{\odot}$/L$_{\odot}$ at the end of the evolution.
This means that dark matter is stripped more efficiently than stars and that the stripping affects even the inner
part of the dwarf where the measurements were done.

To make sure we include a sufficient number of stars from the main body of the dwarf and at the same time
avoid the contamination from the tidal tails for all the
following measurements discussed in this section we fix the maximum radius at $r_{\rm m}=0.5$ kpc.
We note that the results of the measurements do not depend strongly on this choice.
Thus, for each output we select
stars within $r_{\rm m}$, find the principal axes of the stellar component from the tensor of inertia and rotate the
stellar positions and velocities to align them with the principal axes. In the following we will always refer to the
major, intermediate and minor axis of the stellar component as $x$, $y$ and $z$ respectively and the corresponding
axis lengths as $a$, $b$ and $c$. Having aligned the
stellar distribution in this way we estimated the axis ratios as a function of time. The results in terms of $c/a$ and
$b/a$ are shown in the upper panel of Figure~\ref{shape} as the red and blue line, respectively. At the first pericentre
passage the $b/a$ value drops significantly from the initial $b/a=1$ characteristic of the disk, while $c/a$ stays
roughly at the same level. This means that the initial disk transforms into a triaxial stellar component. This triaxial
shape is maintained until the end of the simulation, although both $b/a$ and $c/a$ increase.

\begin{figure*}
\begin{center}
    \leavevmode
    \epsfxsize=5cm
    \epsfbox[0 0 185 200]{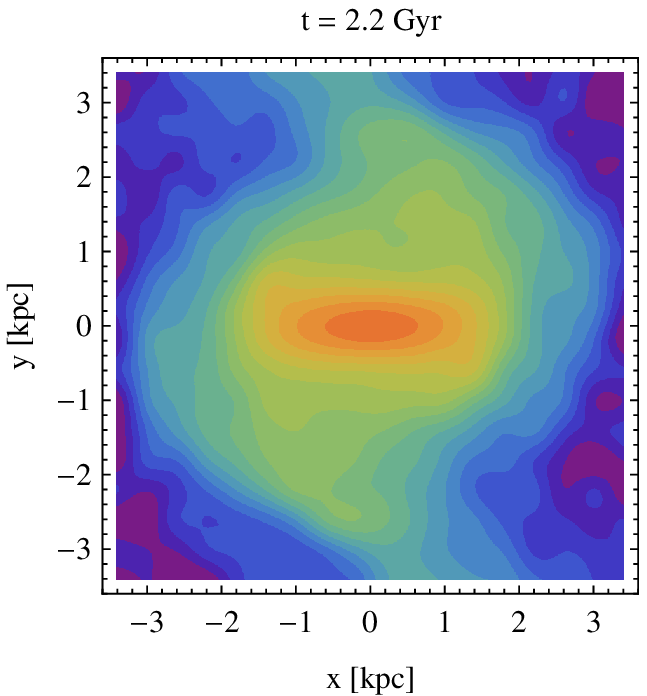}
\leavevmode
    \epsfxsize=5cm
    \epsfbox[0 0 185 200]{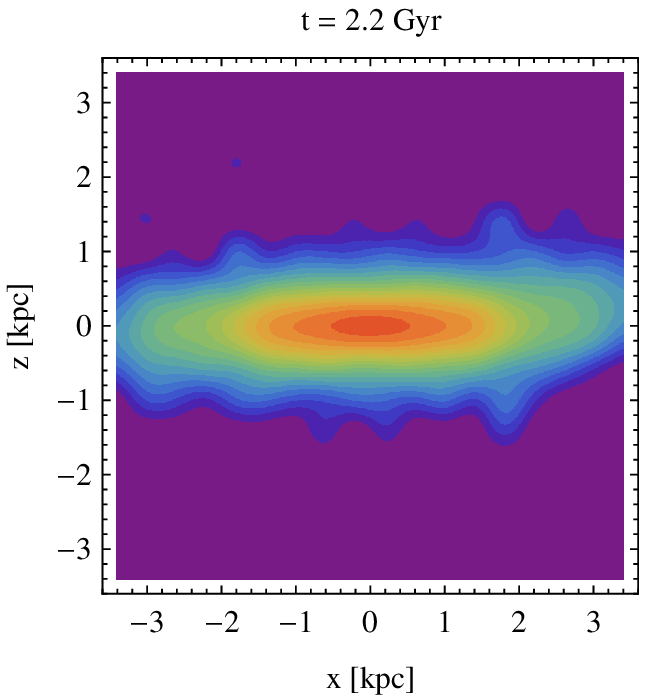}
\leavevmode
    \epsfxsize=5cm
    \epsfbox[0 0 185 200]{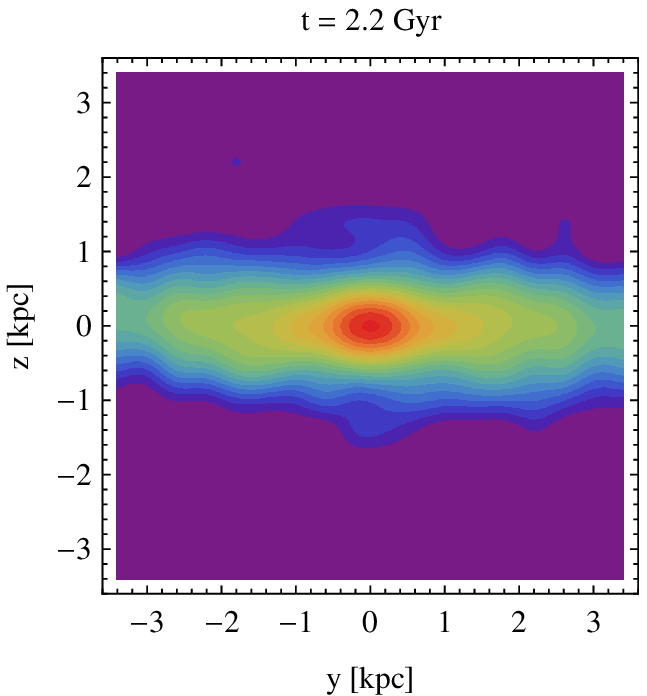}
\leavevmode
    \epsfxsize=0.95cm
    \epsfbox[53 -28 87 149]{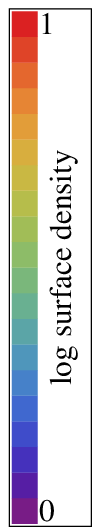}
\leavevmode
    \epsfxsize=5cm
    \epsfbox[0 0 185 200]{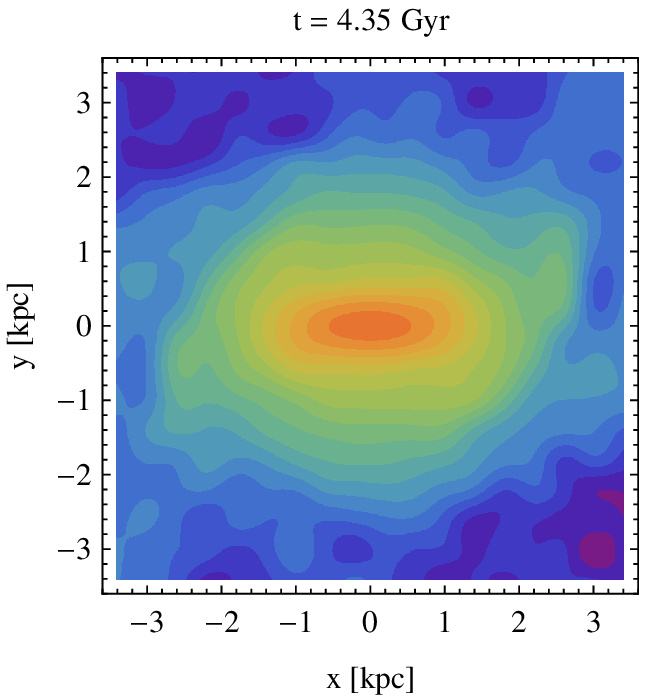}
\leavevmode
    \epsfxsize=5cm
    \epsfbox[0 0 185 200]{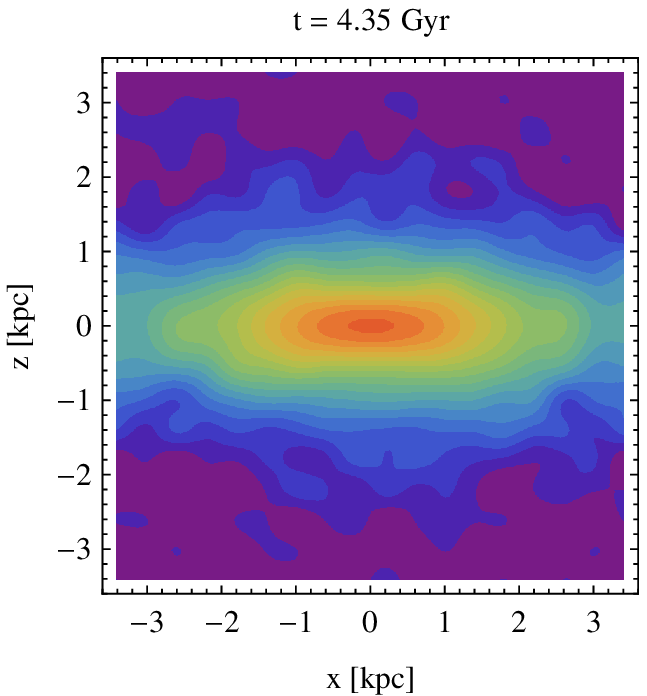}
\leavevmode
    \epsfxsize=5cm
    \epsfbox[0 0 185 200]{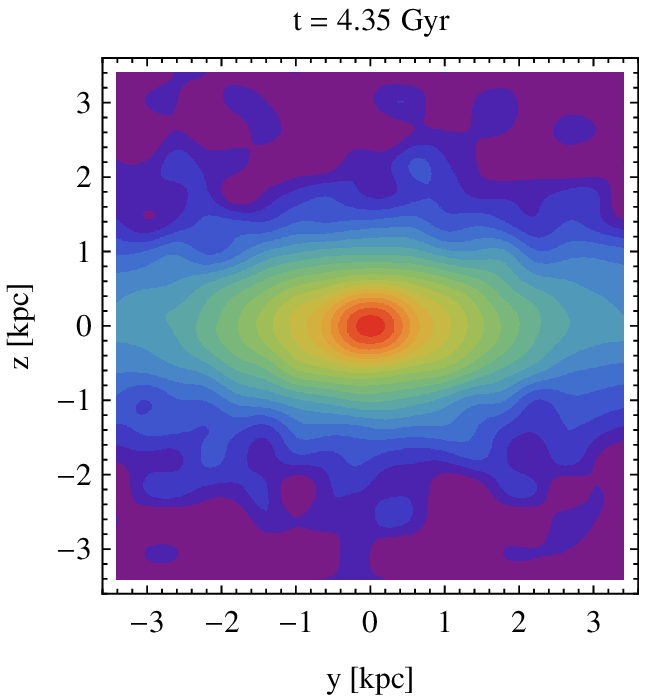}
\leavevmode
    \epsfxsize=0.95cm
    \epsfbox[53 -28 87 149]{legend.eps}
\leavevmode
    \epsfxsize=5cm
    \epsfbox[0 0 185 200]{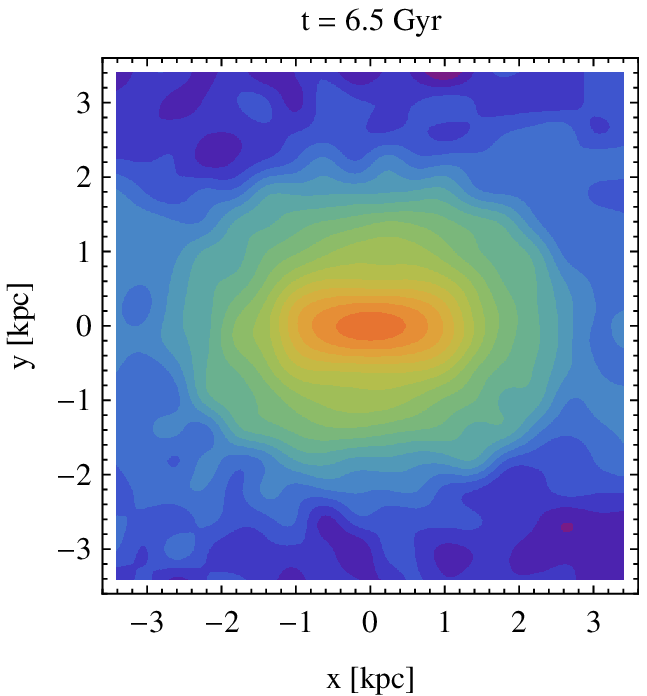}
\leavevmode
    \epsfxsize=5cm
    \epsfbox[0 0 185 200]{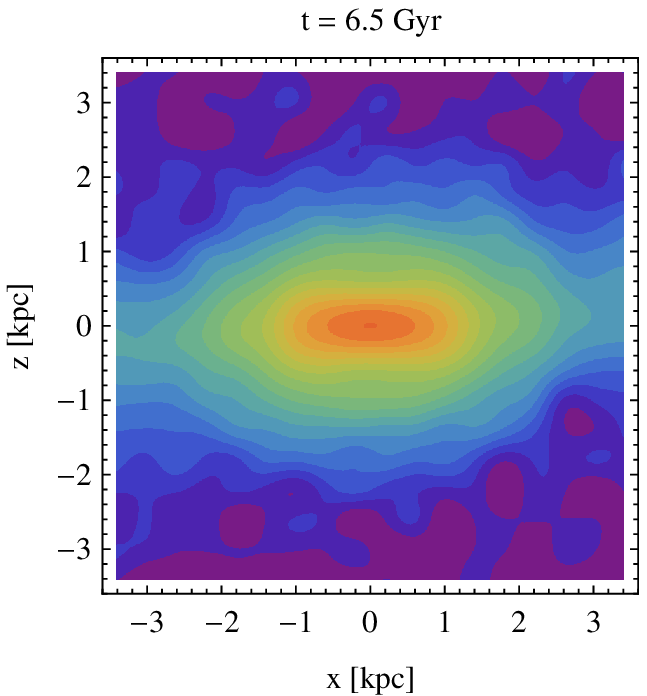}
\leavevmode
    \epsfxsize=5cm
    \epsfbox[0 0 185 200]{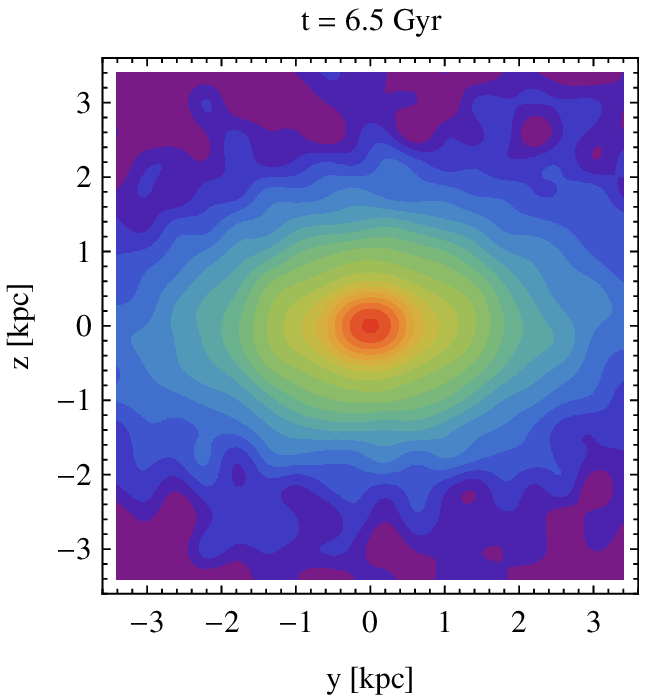}
\leavevmode
    \epsfxsize=0.95cm
    \epsfbox[53 -28 87 149]{legend.eps}
\leavevmode
    \epsfxsize=5cm
    \epsfbox[0 0 185 200]{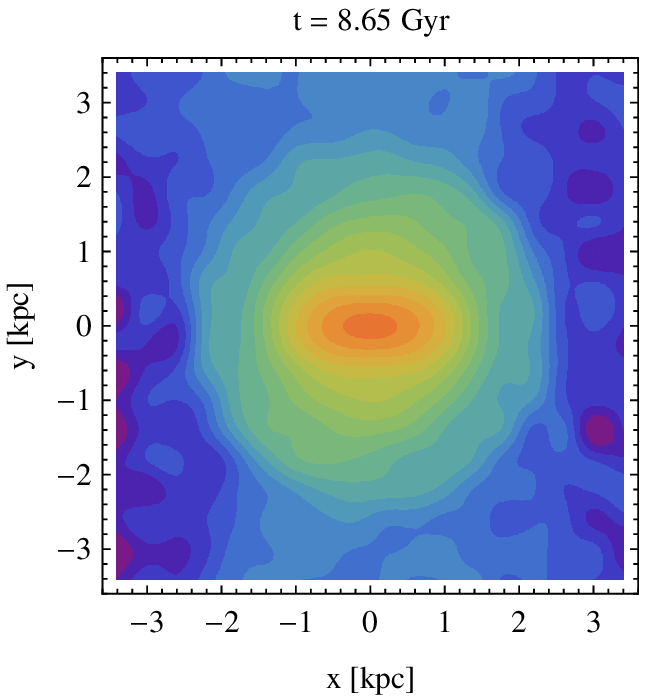}
\leavevmode
    \epsfxsize=5cm
    \epsfbox[0 0 185 200]{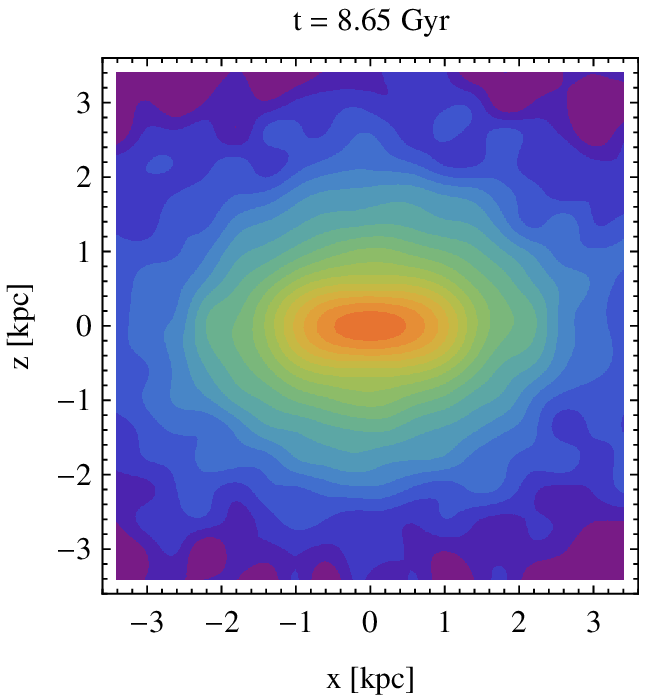}
\leavevmode
    \epsfxsize=5cm
    \epsfbox[0 0 185 200]{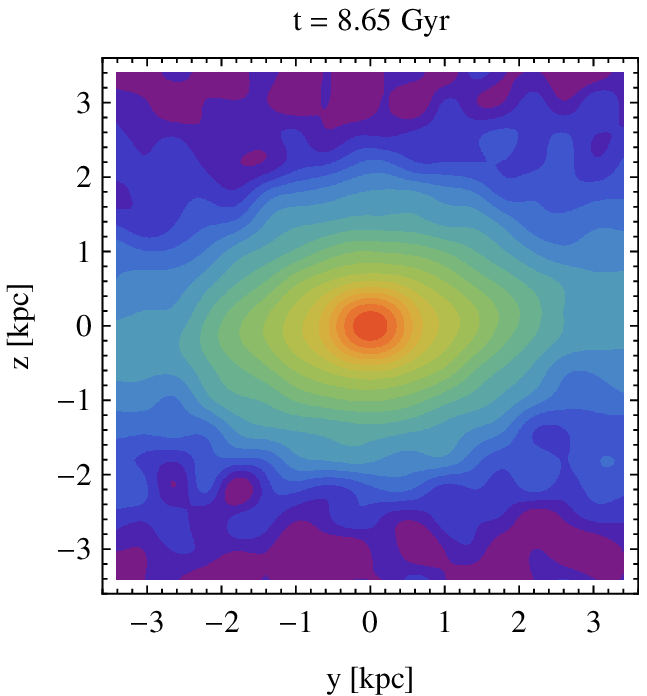}
\leavevmode
    \epsfxsize=0.95cm
    \epsfbox[53 -28 87 149]{legend.eps}
\end{center}
\caption{Surface density distributions of the stars in the dwarf at subsequent apocentres at
$t=2.2, 4.35, 6.5, 8.65$ Gyr (rows) and along different lines of sight: the shortest ($z$), intermediate ($y$)
and longest ($x$) axis of the stellar component (columns, from left to right). The surface density
measurements were normalized to the maximum value $\Sigma_{\rm max} = 9.8 \times 10^5$ stars kpc$^{-2}$ occuring
for the line of sight along the $x$ axis at $t=2.2$ Gyr. Contours are equally spaced in $\log \Sigma$ with
$\Delta \log \Sigma = 0.05$.}
\label{surdenrot}
\end{figure*}

The properties of the shape are further illustrated in the lower panel of Figure~\ref{shape} where the red line shows
the triaxiality parameter $T = [1-(b/a)^2]/[1-(c/a)^2]$. The values of the parameter $0 < T < 1/3$ indicate the oblate
shape, the values $1/3 < T < 2/3$ a triaxial shape and $2/3 < T < 1$ a prolate shape. With $T > 0.7$ almost at all
outputs after the first pericentre we conclude that the shape of the stellar component is decidedly prolate. This
strongly suggests that at the first pericentre passage the stellar component of the dwarf formed a bar. To confirm
this statement we also measured the bar mode $A_2$ from the positions of the stars projected onto the
$xy$ plane (along the shortest axis). In general, the amplitude of the $m$th Fourier mode of the discrete distribution
of stars in the simulated galaxy is calculated as $A_m = (1/N) \left| \Sigma^{N}_{j=1} \exp (i m \phi_j) \right|$
where $\phi_j$ are the particle phases in cylindrical coordinates and $N$ is the total number of stars
(Sellwood \& Athanassoula 1986; Debattista \& Sellwood 2000).
The value of the bar mode $m=2$ is shown with the second (blue) line in the lower panel
of Figure~\ref{shape}. As values of $A_2$ are above 0.2 at all times after the first pericentre we conclude that indeed
a bar is formed at this time and this shape persists until the end of the simulation.

This conclusion is further supported by the measurements of the kinematic properties of the stellar component.
The kinematic measurements were performed using a system of spherical coordinates
such that at every simulation output
the angle $\theta$ measures the angular distance from the shortest axis ($z$) of the
stellar component, while $\phi$ is measured in the $xy$ plane. In the
upper panel of Figure~\ref{kinematics} we plot the velocity dispersions of the stars in the three coordinates. The
three lines from the top to the bottom correspond to $\sigma_r$ (red), $\sigma_{\phi}$ (green) and $\sigma_{\theta}$
(blue) respectively. At the first pericentre passage all dispersions increase, signifying the transition from the
overall streaming motion of the stars (rotation) to the random motions manifesting themselves via the velocity
dispersion. However, the radial dispersion increases most as expected in the case of the formation of a bar which
is supported by more radial orbits. Later on the dispersions decrease due to mass loss, except for $\sigma_{\theta}$
which remains roughly constant as a result of the thickening of the stellar component in time.

The middle panel of Figure~\ref{kinematics} shows the overall contribution of random versus ordered motion in terms of
a 1D velocity dispersion $\sigma = [(\sigma_r^2 + \sigma_{\phi}^2 +  \sigma_{\theta}^2)/3]^{1/2}$ (black line)
and the mean velocity (blue line) around the shortest axis (along the $\phi$ coordinate of the spherical system,
$V = V_{\phi}$) i.e. the rotation velocity. The overall trend is for the rotation
to decrease (especially at the first pericentre passage) and the dispersion to remain roughly constant in time
or slightly decrease due to mass loss.
The transition from mostly circular orbits of the stars (in the initial disk) to more radial orbits in the bar is also
very well visible in the evolution of the anisotropy parameter of the stars
$\beta = 1 - (\sigma_{\theta}^2 + \sigma_{\phi}^{'2})/(2 \sigma_{r}^2)$ where the second velocity moment
$\sigma_{\phi}^{'2} =\sigma_{\phi}^2 + V_{\phi}^2$ includes rotation. The dependence of $\beta$ on time is
shown in the lowest panel of Figure~\ref{kinematics}. Clearly, the circular orbits of stars ($\beta<0$)
in the initial disk are replaced after the first pericentre by more radial orbits ($\beta>0$)
which survive until the end of the simulation.

Figure~\ref{surdenrot} shows the surface density distributions of the stars in the dwarf at subsequent apocentres of the
orbit (from the second at $t=2.2$ Gyr to the fifth at $t=8.65$ Gyr). The rows of the Figure correspond to the different
times of the simulation, while the columns to different lines of sight: along the shortest $z$ (face-on),
intermediate $y$ (side-on) and longest $x$ (end-on) axis of the stellar distribution (as determined
from stars within 0.5 kpc). In the left and middle column the
line of sight is perpendicular to the bar and the bar is clearly visible. At the second apocentre (at $t=2.2$ Gyr) the
stellar component of the dwarf is still restricted to the initial plane of the disk. At later times the stellar component
becomes thicker and at the final apocentre the outer density contours are almost spherical while the
bar-like shape is only preserved in the central part of the dwarf.

\section{The strength of the bar}

In the previous section we measured the global properties of the dwarf galaxy as a function of time, finding convincing
evidence for the formation of a bar after the first pericentre.
The strength of the bar can be quantified in more detail by measuring the profile of the bar mode $A_2$ as a function
of a cylindrical radius $R$. The coordinate system for these measurements was chosen so that $R$ is in the disk
equatorial plane ($xy$) as determined previously from the principal axes of the stellar component within 0.5 kpc.

\begin{figure}
\begin{center}
    \leavevmode
    \epsfxsize=7cm
    \epsfbox[10 10 140 285]{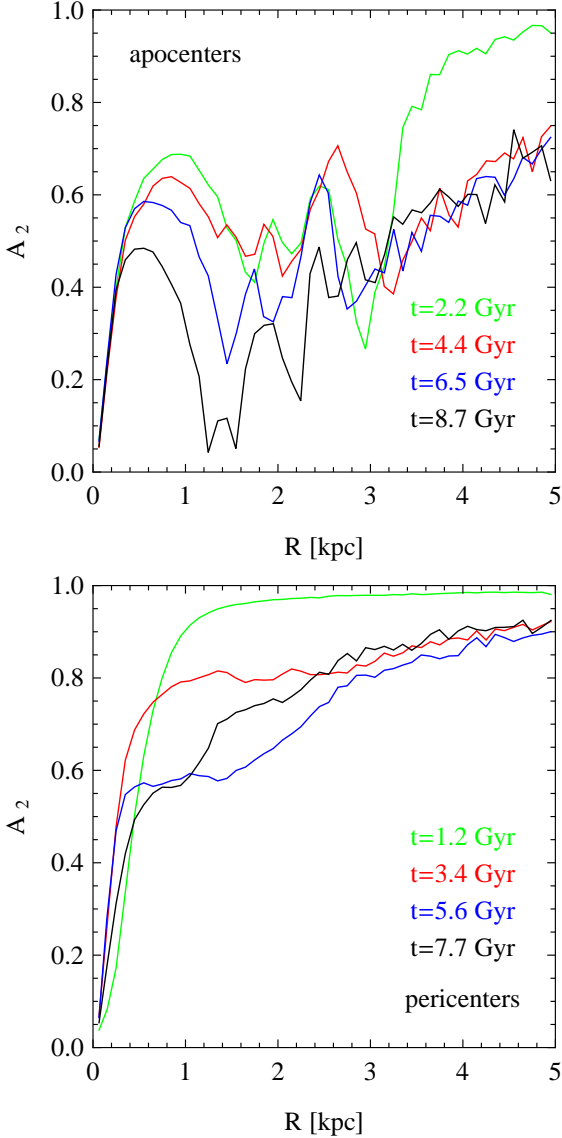}
\end{center}
\caption{The bar mode $A_2$ as a function of cylindrical radius $R$. The upper panel shows the measurements at
subsequent apocentres and the lower panel at subsequent pericentres. At apocentres the bar mode curves show a
characteristic shape with a maximum, while at pericentres they are monotonically increasing.}
\label{a2apoperi}
\end{figure}

The measurements of the values of the bar mode as a function of $R$ are shown in Figure~\ref{a2apoperi}
at subsequent apocentres (upper panel)
and pericentres (lower panel). At pericentres the dwarf is stretched by tidal forces from the Milky Way so that
the bar mode increases monotonically with radius. At apocentres, when the dwarf recovers its equilibrium, the bar mode
displays a characteristic shape, growing with radius up to a maximum value and then decreasing. After reaching a minimum
value, $A_2$ increases again as we transit from the bound component of the dwarf to the tidal tails. The tidal tails
are symmetrical elongated features on both sides of the dwarf (see e.g. {\L}okas et al. 2013)
which obviously results in the $A_2$ approaching unity at large radii. The maximum value of the $A_2$ mode decreases
with time which means that the bar becomes weaker as the evolution proceeds, a feature that was not obvious from the
global single-value measurement of $A_2$ shown in the lower panel of Figure~\ref{shape}.

\begin{figure}
\begin{center}
    \leavevmode
    \epsfxsize=7cm
    \epsfbox[10 10 140 285]{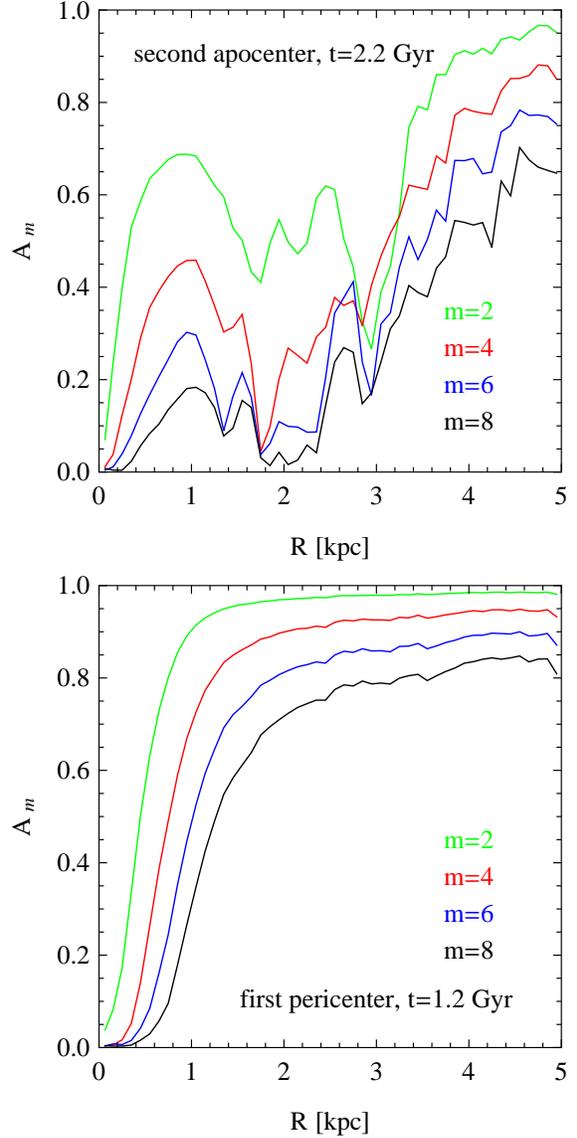}
\end{center}
\caption{The even modes $A_m$ as a function of cylindrical radius $R$. The upper panel shows the profiles of the
first four even modes at the second apocentre ($t=2.2$ Gyr) and the lower panel the analogous measurements at the
first pericentre passage ($t=1.2$ Gyr). The modes all have a similar shape and preserve the hierarchy with
the lower order modes having always higher values.}
\label{amapoperi}
\end{figure}

It is also instructive to look at the higher order Fourier modes of the distribution of the stars. Figure~\ref{amapoperi}
compares the profiles of the non-zero even modes with $m=2,4,6,8$ at the second apocentre (upper panel)
and the first pericentre (lower panel). We note that the odd modes (not shown) all have very low amplitudes
within the main body of the dwarf which means that the distribution of the stars is symmetrical.
The measurements of even modes show that while
the $m>2$ even modes are systematically lower than the most significant $m=2$ bar mode and preserve the hierarchy at
all times, they nevertheless assume values decidedly different from zero. This means that the density distribution in the
tidally induced bar in our simulation cannot be described by the $m=2$ alone but higher even order components are
not negligible. Interestingly, the same behaviour was seen in simulated galaxies forming bars in isolation
by Athanassoula \& Misiriotis (2002, their figure 7) and by Ohta, Hamabe \& Wakamatsu (1990) who studied surface
photometry of six real barred spirals (see their figure 4). Such a hierarchy of modes thus seems to be a general
feature of barred galaxies, independent of their size and of the way they formed.

\begin{figure}
\begin{center}
    \leavevmode
    \epsfxsize=8cm
    \epsfbox[5 5 182 182]{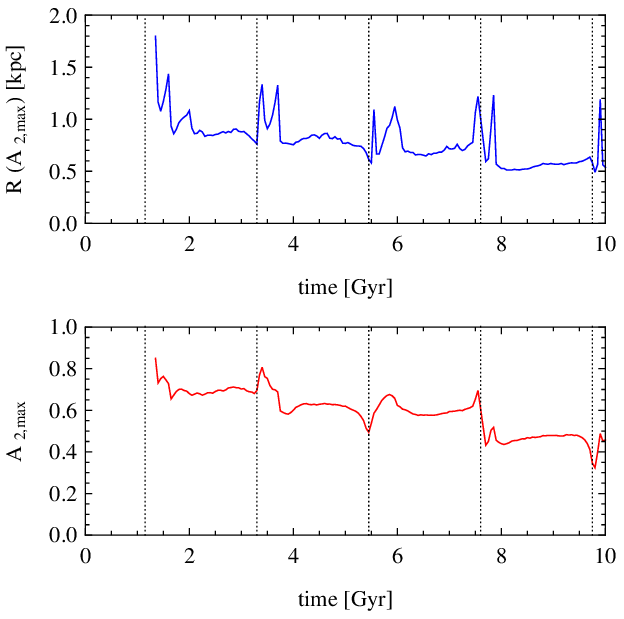}
\end{center}
\caption{The radius at which the first maximum of the bar mode $A_{\rm 2,max}$ occurs (upper panel) and the value
of the maximum of the bar mode $A_{\rm 2,max}$ (lower panel) as a function of time. Vertical dotted lines indicate
pericentre passages.}
\label{a2max}
\end{figure}

We adopt the maximum value of the bar mode $A_{\rm 2,max}$ as the measure of the strength of the bar.
Figure~\ref{a2max} shows in the upper panel the cylindrical radius $R$ at which the first maximum of the bar mode
occurs as a function of time. The measurements are only significant between pericentres since near pericentres
the profile of $A_2$ is increasing and there is no well defined maximum (see Figure~\ref{a2apoperi}).
The lower panel of the Figure plots the
value of the maximum bar mode $A_{\rm 2,max}$ as a function of time. These detailed measurements confirm the
impression from Figure~\ref{a2apoperi}: in the long run the maximum of the bar mode decreases from about 0.7 after
the first pericentre to about 0.45 after the fourth one. Thus we conclude that the bar becomes weaker in time and
the changes in the bar strength are most significant at pericentres while between them $A_{\rm 2,max}$ remains
roughly constant.

\section{The length of the bar}

\begin{figure}
\begin{center}
    \leavevmode
    \epsfxsize=7cm
    \epsfbox[2 0 190 200]{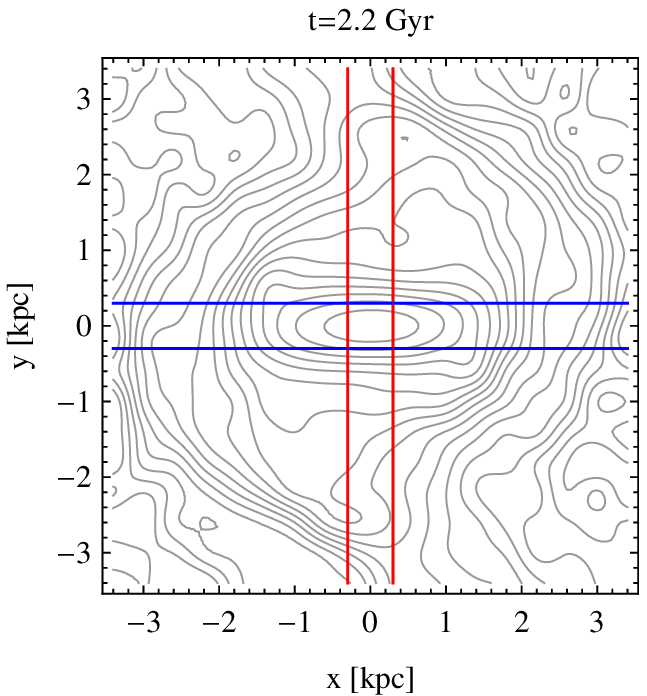}
    \leavevmode
    \epsfxsize=7cm
    \epsfbox[0 0 365 355]{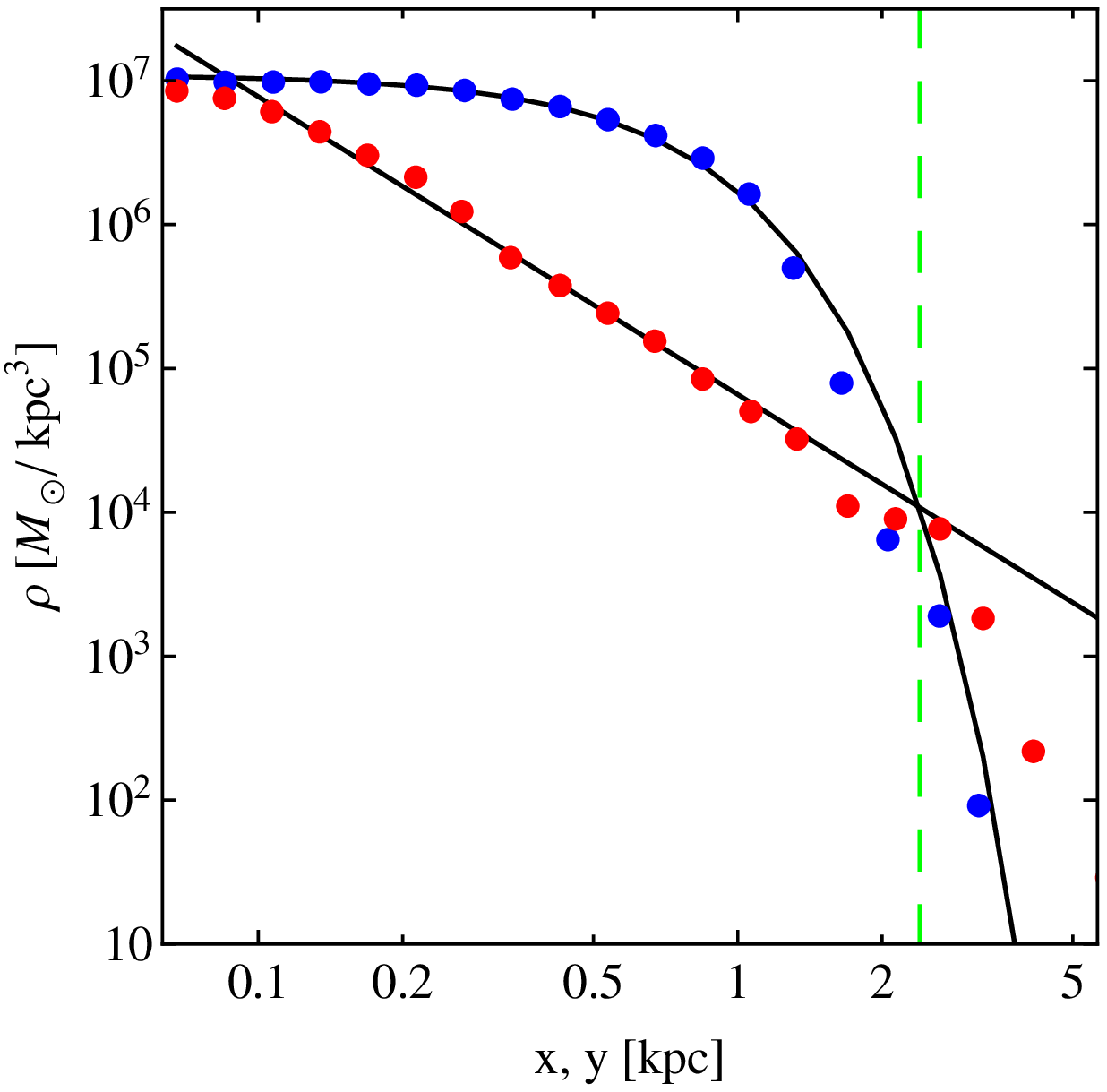}
\end{center}
\caption{Upper panel: an example of the measurement of the density profiles at the second apocentre ($t=2.2$ Gyr).
The contours indicate the levels of equal surface
density distribution of the stars with the bar clearly visible along the $x$ axis. The measurements are done by
counting stars in cylinders of radius 0.3 kpc in bins spaced equally in log $x$ (along the bar, blue lines)
and $y$ (perpendicular to the bar, red lines). Lower panel: density profiles measured for the output shown in the upper
panel. The blue (red) dots indicate measurements along (perpendicular to) the bar. Solid black lines show the
analytic fits to the measured profiles. The vertical green dashed line indicates the radius where the two fitted
profiles are equal, adopted as the measure of the length of the bar. }
\label{density}
\end{figure}

\begin{figure}
\begin{center}
    \leavevmode
    \epsfxsize=8cm
    \epsfbox[5 5 182 270]{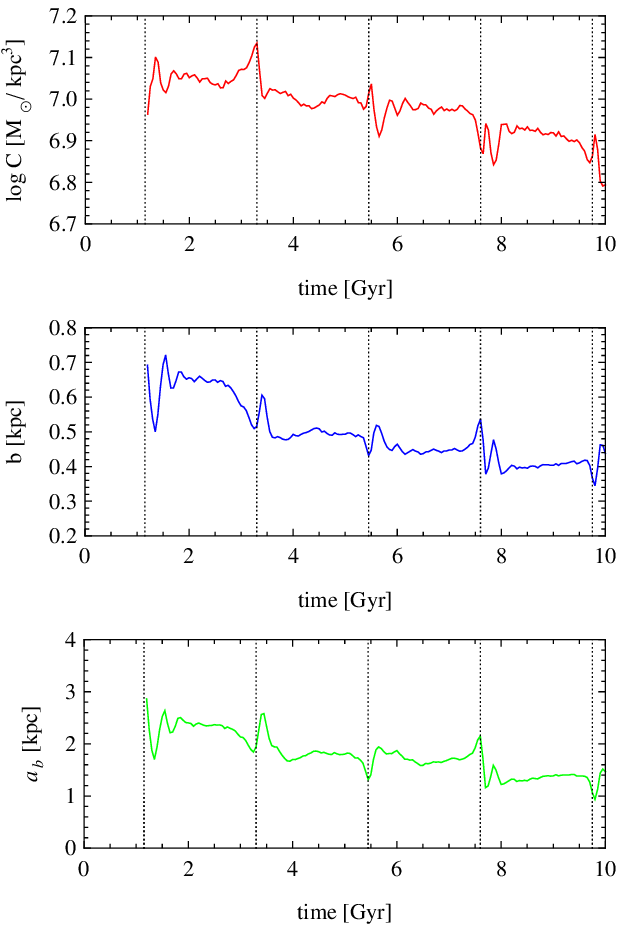}
\end{center}
\caption{The upper and middle panels show the parameters of the formula $\rho_1 (x) = C \exp [- (x/b)^{3/2}]$ fitted
to the measured density profiles along the bar as a function of time: the normalization $C$ and the scale-length $b$.
The lower panel plots the length of the bar $a_{\rm b}$ defined as the radius where the fitted density profiles
along the bar and perpendicular to it converge. Vertical dotted lines indicate
pericentre passages.}
\label{barlength}
\end{figure}

Athanassoula \& Misiriotis (2002) discussed different ways to measure the bar length.
In priciple, the profiles of $A_2$ such as those shown in Figure~\ref{a2apoperi} could be used to determine the
length of the bar. Such a procedure works in general for simulated bars forming in isolated disk galaxies as in such
cases the profile of $A_2$ declines smoothly after reaching the maximum value. One can then find e.g. the radius where
$A_2$ drops to some fraction (e.g. $1/2$) of the maximum value and use this scale as the measure of the bar length.
In our case however, the profile does not always drop sufficiently before it starts to increase again due to
the transition to the tidal tails.
One solution would be to find a minimum of $A_2$ and adopt the corresponding scale as the length of the bar. However,
even after smoothing our $A_2$ profiles are rather noisy and such measurements result in very imprecise estimates of
the bar length.

We therefore measured the length of the bar using a different approach, similar to method (v) discussed by
Athanassoula \& Misiriotis (2002, section 8). Namely, we calculated the density of stars
along the bar major axis (i.e. along the $x$ axis of the stellar component)
and perpendicular to it (along the $y$ axis)
in cylinders of radius 0.3 kpc and logarithmic bins in $x$ and $y$. An example of such measurements after $t = 2.2$ Gyr
from the start of the simulation (second apocentre) is shown in the
upper panel of Figure~\ref{density}. The contours indicate equal levels of surface density of stars similar to those
plotted in Figure~\ref{surdenrot} with the bar well visible along the $x$ axis in the inner part of the picture. The
cylinders of stars selected for the measurements are indicated with blue (along the bar) and red (perpendicular to the
bar) lines. The measured density profiles are plotted as dots of the corresponding colours in the lower panel of
Figure~\ref{density}. As expected, the measurement along the bar (blue points) is rather flat in the centre,
reflecting the approximately constant density distribution of the stars in the bar. The difference between the densities
measured in the two directions first grows with radius and then the two converge. We will adopt the scale where the
two densities converge as the measurement of the length of the bar.

In order to avoid the noise due to the limited number of stars in the outer bins we fitted the profiles with analytic
functions. We find that the density along the bar is well approximated at all times by an exponential
$\rho_1 (x) = C \exp [- (x/b)^{3/2}]$ where $C$ is the normalization constant and $b$ is the characteristic
scale-length. The density distribution perpendicular to the bar is well approximated by a simple power law
$\rho_2 (y) = D y^{-d}$. For every simulation output we measured the density profiles in this way and fitted the
formulae to the measurements. The fitted values of the $C$ and $b$ parameters of the density distribution along
the bar are especially interesting and their evolution is shown as a function of time in the upper and middle panel of
Figure~\ref{barlength}. We see that the values of both parameters decrease in time reflecting the stripping of the
stars (decrease of normalization) and shortening of the bar (decrease of scale-length). The parameters of the power-law
fit $\rho_2 (y)$ do not show any clear trend in time. Both the normalization $D$ and the power-law index $d$ stay
roughly constant in time with $d$ in the range of 2-2.5.
We note that the particular choice of the fitting formulae does not have to apply to other kinds of
bars. We used them mainly as a tool to smooth the results of bar length measurements and because the formulae
were general enough to accommodate the density profiles of our bar at all times.

Solving $\rho_1 (x) = \rho_2 (y)$ with the fitted parameters we find for each simulation output the scale at which
both density profiles converge. The scale, which we identify with the length of the bar, $a_{\rm b}$, is plotted
as a function of time in the lower panel of Figure~\ref{barlength}. The length of the bar thus decreases during the
evolution from $a_{\rm b} = 2.4$ kpc after the first pericentre to $a_{\rm b} = 1.3$ kpc after the fourth. Note that
the decrease of the bar length occurs mainly at pericentre passages so it is due to tidal shocking and not to
secular evolution since between pericentres the length remains approximately constant in time.

\section{The pattern speed of the bar}

\begin{figure}
\begin{center}
    \leavevmode
    \epsfxsize=8.1cm
    \epsfbox[5 2 182 267]{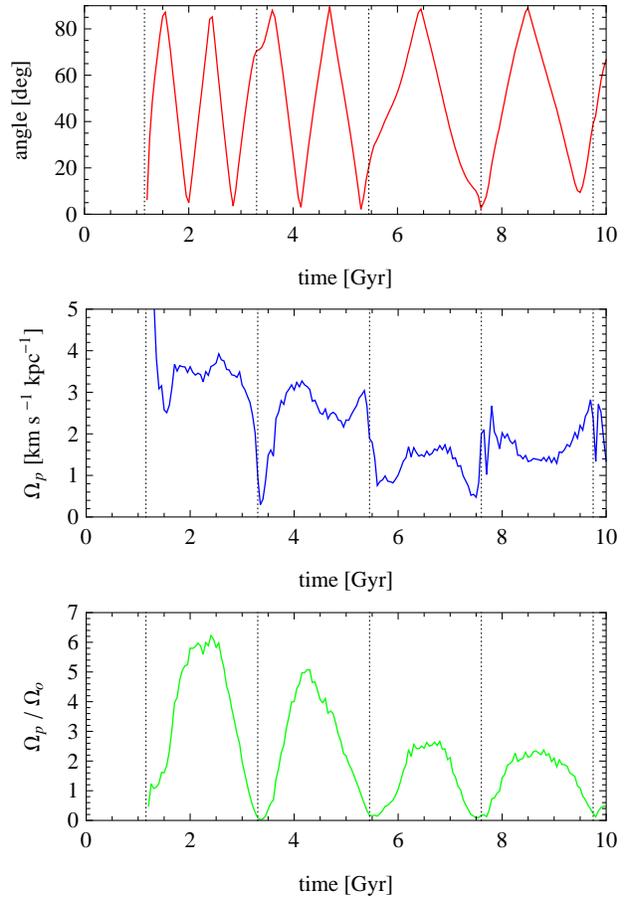}
\end{center}
\caption{Upper panel: the angle between the nearest side of the bar's major axis and a fixed axis of the
simulation box as a function of time. Middle panel: the pattern speed of the bar as a function of time. Lower panel:
the ratio of the bar's pattern speed to the angular velocity of the dwarf on the orbit.
Vertical dotted lines indicate pericentre passages.}
\label{patternspeed}
\end{figure}

We finally also look at the pattern speed of the bar.
The simplest way to visualize its variability is to measure the angle
between the major axis of the stellar component and a fixed axis of the simulation box. For this measurement we use the
orientation of the major axis determined as before using stars within a constant radius of 0.5 kpc. The angle
between the major axis of the bar and a fixed axis in the initial orbital plane of the dwarf is shown as a function
of time in the upper panel of Figure~\ref{patternspeed}. At each output we calculated the angle between the fixed
axis and the nearest part of the bar, which is why the angle is always in the range between 0 and 90 degrees.
The measurements start after the first pericentre when the bar is formed. The bar lies almost exactly in
the orbital plane of the dwarf at all times as indicated by the values of the angle covering the whole range
of 0-90 degrees. The tumbling of the bar seems much slower in the second half of the evolution (after the third
pericentre passage), as shown by the much slower changes of the angle.

The actual pattern speed calculated as a change between the directions of the bar's major axis in subsequent
simulation outputs is plotted in the middle panel of Figure~\ref{patternspeed}. The pattern speed decreases in
the long run but shows a rather strong variability, mostly at the pericentre passages, but not only. We note
that the variability of the pattern speed mirrors closely our measurements of the mean rotation velocity
of the stars discussed in section 3 (see the blue line in the middle panel of Figure~\ref{kinematics}). This means
that after the first pericentre the rotation is mostly due to the tumbling of the bar.

\begin{figure}
\begin{center}
    \leavevmode
    \epsfxsize=7cm
    \epsfbox[0 0 185 205]{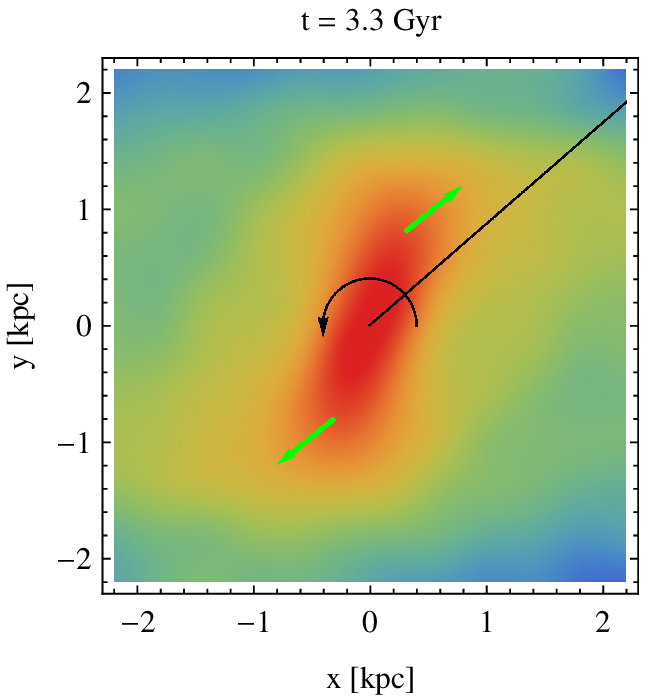}
\leavevmode
    \epsfxsize=7cm
    \epsfbox[0 0 185 205]{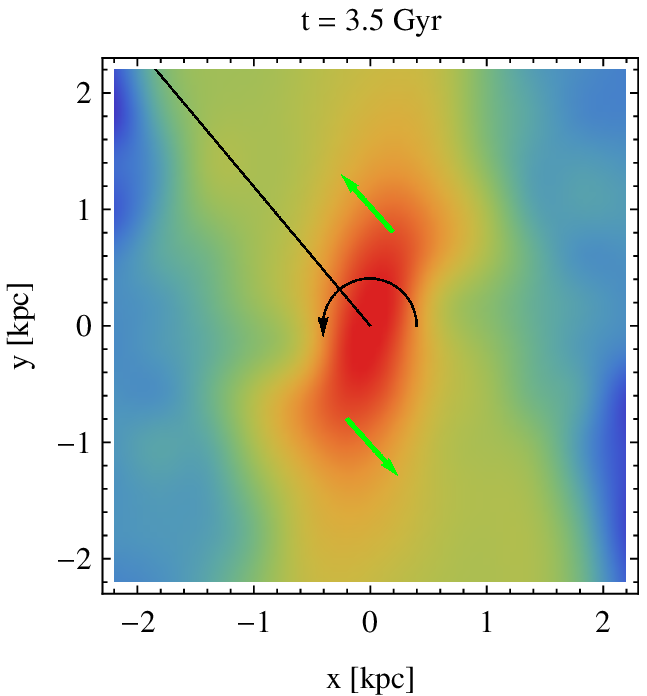}
\end{center}
\caption{The change of direction of tidal torque from the Milky Way near the second pericentre passage. Plots show
the surface distribution of the stars in the dwarf projected onto the orbital plane at $t=3.3$ (upper panel)
and $t=3.5$ Gyr (lower panel). The colour coding is similar as in Figure~\ref{surdenrot} but
normalized to the central density value for this stage and line of sight
$\Sigma_{\rm max} = 6.8 \times 10^5$ stars kpc$^{-2}$.
In both panels the curved arrows in the centre indicate the direction of rotation of the
bar. The solid lines show the direction towards the Milky Way. Green arrows indicate the tidal forces acting on the bar.
Between the outputs the direction of the tidal torque changes from one slowing down the bar to one speeding it up.}
\label{surdenperi}
\end{figure}

A particularly interesting behaviour occurs at the second pericentre passage when the pattern speed drops
almost to zero, i.e. the bar almost stops rotating but then speeds up again.
The variation of the pattern speed at this moment (and its general evolution) can be understood by referring to
Figure~\ref{surdenperi} where we show the orientation of the bar with respect to the direction towards the centre
of the Milky Way. The upper and lower panel show respectively the
projections of the stellar component onto the orbital plane at $t=3.3$ and $t=3.5$ Gyr, that is just before and
just after the second pericentre. The circular arrow in the centres marks the anti-clockwise direction of the bar's
rotation. In each panel the solid black line indicates the direction towards the Milky Way
and the two green arrows show the direction of tidal forces acting on the two sides of the bar. At the earlier time
depicted in the Figure (upper panel) the torque due to the tidal forces is directed so that it slows the bar.
At the later time (lower panel) the orientation of the bar with respect to the direction to the Milky Way changes
and the torque now speeds up the bar. The result is for the bar to regain the pattern speed up to almost the same
level as before the pericentre.

The subsequent changes of the pattern speed, the more violent ones at pericentres, as well as the milder ones
between pericentres can all be traced to a particular orientation of the bar with respect to the tidal force
acting at a given time. In particular, at the third pericentre passage the bar is systematically slowed down
(the orientation is similar as in the upper panel of Figure~\ref{surdenperi}), while at the fourth pericentre
the bar is continuously accelerated (the orientation is similar as in the lower panel of Figure~\ref{surdenperi}).

It is also interesting to compare the pattern speed of the bar $\Omega_{\rm p}$ to the angular velocity
of the dwarf galaxy on its orbit $\Omega_{\rm o}$.
The ratio of the two quantities is plotted in the lower panel of Figure~\ref{patternspeed}. At pericentres
the dwarf obviously moves very fast on its rather eccentric orbit so the ratio $\Omega_{\rm p}/\Omega_{\rm o}$
is close to zero. Between pericentres the ratio increases up to a factor of a few and at apocentres is never below two.
Note that $\Omega_{\rm p}/\Omega_{\rm o} =1$ would mean that the bar is tidally locked as the Moon is locked to the
Earth and only one and always the same side of the bar would be directed towards the Milky Way. This is clearly not the
case. However, an interesting evolutionary stage takes place near the fourth apocentre of the orbit ($t=6-7$ Gyr)
when $\Omega_{\rm p}/\Omega_{\rm o} \approx 5/2$ and the angle between the bar and the direction to the Milky Way
stays in the range of $0-20$ degrees. This points to a possibility of a 5/2 resonance between the rotational and orbital
motion of the bar, similar to the 3/2 resonance of Mercury around the Sun (Correia \& Laskar 2004).

\begin{figure}
\begin{center}
    \leavevmode
    \epsfxsize=7cm
    \epsfbox[0 0 190 190]{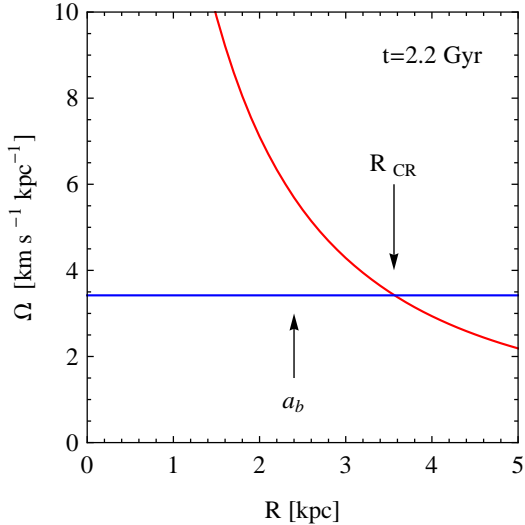}
\end{center}
\caption{The circular frequency of the dwarf galaxy as a function of radius (red line) in comparison with the
pattern speed of the bar $\Omega_{\rm p}$ (blue line) at the second apocentre ($t=2.2$ Gyr).
These two quantities are equal at the corotation radius $R_{\rm CR} = 3.6$ kpc marked with a black arrow.
The second black arrow
indicates the length of the bar at this time $a_{\rm b} = 2.4$ kpc. The ratio of the two $s = R_{\rm CR}/a_{\rm b}=1.5$
is close to unity, so the bar is rather fast.}
\label{corotationapo}
\end{figure}

Finally, we estimate the speed of the bar in terms of the quantity $s = R_{\rm CR}/a_{\rm b}$ where $R_{\rm CR}$
is the corotation radius (where the circular frequency of the dwarf galaxy $\Omega$ equals
the pattern speed $\Omega_{\rm p}$) and $a_{\rm b}$ is the length of the bar estimated in the previous section.
According to theory of bars (Binney \& Tremaine 2008) they can only exist at radii $R<R_{\rm CR}$ so all bars
have $s > 1$ and are classified as fast if $s \approx 1$ or slow when $s \gg 1$. We calculated the corotation radii
for a number of outputs and compared them with corresponding bar lengths. An example of such a comparison
for the second apocentre ($t=2.2$ Gyr) is shown in Figure~\ref{corotationapo}. The red line is the circular frequency
of the dwarf galaxy $\Omega = [G M(r)/r^3]^{1/2}$ as a function of radius (calculated with radial step of 0.1 kpc)
and the blue line is the pattern speed from the middle panel
of Figure~\ref{patternspeed}. The two are equal at $R_{\rm CR} = 3.6$ kpc while the bar length at this time is
$a_{\rm b} = 2.4$ kpc (see the lower panel of Figure~\ref{barlength}). This gives us $s = 1.5$, a value not very different
from unity, thus the bar at this stage may be classified as fast.

\begin{table}
\begin{center}
\caption{Estimates of the speed of the bar at subsequent apocentres. }
\begin{tabular}{clccc}
apocentre & time  & $R_{\rm CR}$ & $a_{\rm b}$ & $R_{\rm CR}/a_{\rm b}$ \\
          & [Gyr] &  [kpc]       & [kpc]       &                        \\
\hline
2        &  2.2    &   3.6    &   2.4   & 1.5  \\
3        &  4.35   &   3.4    &   1.8   & 1.9  \\
4        &  6.5    &   4.7    &   1.6   & 3.0  \\
5        &  8.65   &   4.4    &   1.3   & 3.3  \\
\hline
\label{tablespeed}
\end{tabular}
\end{center}
\end{table}

Similar calculations at all apocentres give the results listed in Table~\ref{tablespeed}. We restricted ourselves to
the measurements at apocentres because near pericentres the pattern speed varies strongly and in particular can be very
low which leads to high and thus meaningless estimates of the corotation radius. We see that at subsequent apocentres
the values of $s$ systematically increase, up to $s=3.3$ at the last apocentre, so the bar becomes slower with time.
This result is not obvious at first sight because $R_{\rm CR}$ and $a_{\rm b}$ behave differently in time.
While the bar length $a_{\rm b}$ decreases monotonically between apocentres (see Table~\ref{tablespeed}), the
$R_{\rm CR}$ increases or decreases with time. The latter is itself a combination of the
circular frequency, which decreases monotonically with time
as a result of mass loss, and the pattern speed, which does not have an obvious monotonic behaviour. The overall measure
of the bar speed in terms of $R_{\rm CR}/a_{\rm b}$ confirms however the general impression from the behaviour of the much
simpler quantity such as the mean rotation velocity (see the middle panel of Figure~\ref{kinematics}) or the mean
angular momentum which behaves in a way very similar to the rotation velocity.

The main physical reason for the slow-down of the bar over large
time scales can be traced to the effect of tidal stripping of the stars which happens preferentially to stars on
prograde orbits (as is the case in our simulation) due to resonances between the orbital and intrinsic motion
of the stars (D'Onghia et al. 2010). The stripped stars feed the tidal tails of the dwarf galaxy and
move on their own orbits around the Milky Way. Thus the angular momentum of the stars is taken by the stripped stars
and not transferred to the dark matter particles of the dwarf's halo as the angular momentum of the halo
does not significantly change during the evolution.

\section{Discussion}

We studied the formation and evolution of a stellar bar induced in a disky dwarf galaxy orbiting the Milky Way
by tidal forces from the host. We measured the main properties of the bar such as its strength, length and pattern
speed as a function of time and related the pattern speed to the dwarf's circular frequency.
The comparison between the two quantities led us to conclude that
while the bar is quite fast at its birth after the first pericentre passage, it becomes slower with time. This has
important consequences for understanding the process of formation of dwarf spheroidal galaxies of the Local Group.

As the tidal evolution proceeds, the bar becomes shorter and thicker and the stellar component changes its shape
towards spherical. One could attempt to explain the shortening of the bar as due to the mass loss that results
in decreasing the circular frequency of the dwarf. If the pattern speed of the bar remained constant in time,
the corotation radius would also decrease. Since the bar cannot extend beyond the corotation radius, this would explain
why it becomes shorter and thus relate the mass loss to the morphological transformation.
The actual behaviour, however, turns out to be more complicated. As we demonstrated in the previous sections,
the pattern speed of the bar is not constant but subject to abrupt changes near pericentres and
more benign ones in the other parts of the orbit, depending on the orientation of the bar major axis with respect
to the direction of the tidal force. This causes the corotation radius to vary non-monotonically with
time. There is thus no simple relation between the length of the bar and the corotation radius. The bar seems to
become shorter just due to randomization of stellar orbits resulting from tidal shocking.

It seems to be generally believed that a common intermediate stage of the evolution of disky dwarfs towards a more
spherical shape is that of bar buckling. Buckling seemed to occur in a large fraction of the cases of tidally
induced bars studied by Mayer et al. (2001). They claimed that buckling contributes to the heating of the disk
even more than the tidal heating itself. The occurrence of buckling is usually accompanied by an increase of the ratio of
velocity dispersions along the shortest axis and in the bar plane $\sigma_z/\sigma_R$ and by non-zero amplitude of the
odd Fourier mode $A_1$ in the edge-on view (e.g. Martinez-Valpuesta et al. 2006).
We have looked for signatures of bending instabilities in our bar by measuring the ratio
$\sigma_z/\sigma_R$ as a function of time
(for stars within radius of 1 kpc). This ratio is close to 0.4 after the first pericentre when the bar forms
and increases steadily with time to about 0.6 at the end of the evolution. There is no abrupt increase of
$\sigma_z/\sigma_R$ that would signify the presence of buckling.
We have also measured the $A_1$ mode for the stars seen along the intermediate axis (as in the middle
column of Figure~\ref{surdenrot}) and did not find it significantly different from zero.

The actual presence of
buckling was only detected by visual inspection of surface density maps such as those in Figure ~\ref{surdenrot}.
Slight asymmetries in the distribution of stars with respect to the bar plane in the edge-on view were observed
for a brief period between 3.5 and 3.8 Gyr from the start of the simulation, that is soon after the second pericentre
passage. As expected, the occurrence of buckling was
accompanied by the decrease in the bar strength ($A_2$ and $A_{2,{\rm max}}$) seen in the lower panels
of Figures~\ref{shape} and \ref{a2max} at these times. This brief period of buckling instability was followed by the
formation of the boxy/peanut shape visible in the edge-on view of the surface distribution of stars at $t=4.35$ Gyr
in Figure~\ref{surdenrot}.

In this paper, we explored the properties of the tidally induced bar only in one initial and rather special configuration,
that of coplanar disks of the dwarf and the host galaxy and the dwarf's disk rotation exactly prograde with respect
to the orbital motion. While the orientation of the Milky Way disk with respect to the orbital plane of the dwarf seems
of little consequence, the angle between the dwarf disk's angular momentum and the orbital angular momentum has
dramatic consequences. Preliminary simulations show that if the dwarf disk orientation is exactly retrograde the
bar does not form at all and the dwarf's stellar component remains disky. For intermediate orientations the bar does form
but it is typically weaker than in the case studied here. The dependence of the properties of tidally induced bars
on this and other parameters will be discussed in follow-up papers.

\section*{Acknowledgements}

This research was supported in part by PL-Grid Infrastructure,
by the Polish National Science Centre under grants NN203580940,
2013/10/A/ST9/00023 and the Polish-French HECOLS collaboration including grant 2013/08/M/ST9/00664.
EA acknowledges financial support to the DAGAL network from the People Programme (Marie Curie Actions)
of the European Union's Seventh Framework Programme FP7/2007-2013/ under REA grant agreement number
PITN-GA-2011-289313. She also acknowledges financial support from the CNES
(Centre National d'Etudes Spatiales - France). We are grateful to the organizers and participants of the
conference ``The Role of Bars in Galaxy Evolution" in Granada in May 2013 for inspiration and discussions.
We would like to thank L. Widrow for providing procedures to generate $N$-body models
of galaxies for initial conditions. EL{\L} is grateful for the hospitality of Laboratoire d'Astrophysique
de Marseille at the time of her visit and
AdP for the hospitality of the Copernicus Center in Warsaw during his visit.
MS, GG and KK acknowledge the summer student program of the Copernicus Center.

\end{document}